\documentclass[prd,twocolumn,floats,floatfix,showpacs,nofootinbib]{revtex4}
\usepackage{graphicx}
\usepackage{dcolumn}
\usepackage{bm}
\usepackage{graphics}
\usepackage{slashed}
\usepackage{amssymb}
\usepackage{natbib}
\usepackage{amsmath}
\usepackage{amsthm}
\usepackage{url}
\usepackage{color}
\newcommand{\fracc}[2]{\frac{\textstyle{#1}}{\textstyle{#2}}}

\begin{document}

\title{Analogue black holes for light rays in static dielectrics}

\author{E. Bittencourt$^{1,2}$} \email{eduardo.bittencourt@icranet.org}
\author{V. A. De Lorenci$^{3,4}$} \email{delorenci@unifei.edu.br}
\author{R. Klippert$^5$} \email{klippert@unifei.edu.br}
\author{M. Novello$^1$}\email{novello@cbpf.br}
\author{J. M. Salim$^1$} \email{jsalim@cbpf.br}

\affiliation{$^1$Instituto de Cosmologia Relatividade Astrofisica ICRA - CBPF,\\
Rua Dr. Xavier Sigaud 150 Urca, 22290-180, Rio de Janeiro, Brazil}
\affiliation{$^2$Sapienza Universit\`a di Roma - Dipartimento de Fisica\\
P.le Aldo Moro 5 - 00185 Rome - Italy and\\
ICRANet, Piazza della Repubblica 10 - 65122 Pescara - Italy}
\affiliation{$^3$Instituto de F\' {\i}sica e Qu\'{\i}mica, Universidade Federal de Itajub\'a, Itajub\'a, MG 37500-903, Brazil}
\affiliation{$^4$ Institute of Cosmology, Department of Physics and Astronomy,
Tufts University, Medford, MA 02155, USA}
\affiliation{$^5$Instituto de Matem\'atica e Computa\c{c}\~ao, Universidade Federal de Itajub\'a, Itajub\'a, MG 37500-903, Brazil}

\pacs{04.20.Cv; 04.20.-q; 11.10.Lm; 42.15.Dp.}

\date{\today}

\begin{abstract}
Propagation of light in nonlinear materials is here studied in the regime of the geometrical optics. It is shown that a spherically symmetric medium at rest with some specific dielectric properties can be used to produce an exact analogue model for a class of space-times which includes spherically symmetric and static black hole solutions. The optical model here presented can be a useful tool to reproduce in laboratory the behavior of optical null geodesics near a compact object with an observable gravitational Schwarzschild radius.
\end{abstract}

\maketitle

\section{Introduction}
Analogue models of general relativity have long ago been considered in the literature. This theme was introduced by Gordon in 1923 \cite{gordon} when an effective geometric interpretation for light propagation in homogeneous dielectrics was proposed. In the following six decades few works on this theme were published (for a review, see Refs. \cite{plebanski1960,landau1971,felice1971,schenberg1971}). However, this research area was warmed up in the 80's with the proposal of a hydrodynamic analogue model of a black hole by Unruh \cite{unruh1981}. Since then, lot of other models were proposed and their consequences investigated. For a review, see Ref. \cite{visser2011} and references therein. The main  expectation is that some tiny effects predicted to occur in the realm of semiclassical gravity could be tested in such analogous to general relativity systems. Of particular interest is the issue of Hawking radiation, which is expected to occur whenever an event horizon is brought forth in a physical system, although some authors have different opinion \cite{belinski}. Such effect was unveiled by Hawking in 1975 \cite{hawking1975}  as a result of the quantization of fields in the spacetime of a black hole. This issue was recently examined in laboratory by means of optic \cite{belgiorno2010} and hydrodynamic \cite{unruh2011} gravitational analogue systems. In both cases Hawking-like radiation was reported to occur.

In the context of electrodynamics the most simple analogue model describing the propagation of light in a continuum media arises by considering constant dielectric coefficients $\epsilon$ and $\mu$, here called electric permittivity and magnetic permeability, respectively. In such case, it can be shown that the paths of light are geodesic in an effective geometry given by
\begin{equation}
g^{\mu\nu} = \eta^{\mu\nu}
-  (1-\mu\epsilon)v^\mu v^\nu
\label{gordon},
\end{equation}
where $\eta^{\mu\nu}$ is the Minkowski metric and $v^\alpha$ is the observer's geodesic four-velocity (assumed to be normalized to unity by simplicity) with respect to which the material medium is at rest. The above metric is known as the Gordon metric \cite{gordon,edu2012}. More general results can be obtained by relaxing the constance of the dielectric coefficients. In this case several models describing different solutions of general relativity were obtained. For instance, almost axially symmetric optical analogues of Schwarzschild black holes are already known \cite{novello4,Yuri,bjp}. For these models, the occurrence of an event horizon is fundamentally dependent upon the vortical motion of the dielectric fluid. A model that does not involve mechanical motion of the medium was already presented \cite{narimanov} in the context of metamaterials, where an effective black hole solution was proposed. In this paper, we investigate further the possibility of producing a material medium at rest with specific dielectric properties in such way that an exact analogue model for Schwarzschild geometry becomes available. Particularly, the behavior of light rays propagating in this medium is studied. We focus on the theoretical aspects of the phenomenon. Experimental verification is already known in some tailored media \cite{sheng,yin}.

The outline of the paper is as follows. In Section \ref{Gordon}, basic aspects of light propagation is reviewed, including the derivation of the dispersion relations for nonlinear material media and the corresponding effective optical metric interpretation. An optical model is presented in Sec.\ \ref{Analogue}. Some kinematical aspects of the propagating optical wave modes in such model are discussed in Sec.\ \ref{Kinematics}, which justify calling it an optical analogue model of spherically symmetric static black hole solutions. Concluding remarks are presented in Sec.\ \ref{Conclusion}, including a brief discussion about the issue of Hawking thermal radiation in this model.
We assume throughout this work that the physics takes place in the flat Minkowski space-time of special relativity, which is described in a general coordinate system with metric $\gamma_{\mu\nu}$. Geometrical units are chosen such that the speed of light in vacuum is $c=1$. We define the completely skew-symmetric pseudo-tensor $\eta^{\alpha\beta\mu\nu}$
such that $\eta^{0123}=1$ when written in Cartesian coordinates.

\section{The generalized Gordon optical metric}
\label{Gordon}
We briefly recall here the main steps \cite{local} to achieve the optical metric description of the wave propagation in material media in the limit of geometrical optics. Let $v^\alpha$ be the observer's geodesic four-velocity (assumed to be normalized to unity for simplicity), with respect to which the material medium is at rest. Maxwell equations inside this medium in Minkowski space-time can be written in covariant notation as
\begin{eqnarray}
P^{\alpha\beta}{}_{;\beta}=J^\alpha,\label{max_diel1}\\[2ex]
({}^*F)^{\alpha\beta}{}_{;\beta}=0,\label{max_diel2}
\end{eqnarray}
where $({}^*F)^{\alpha\beta}=\eta^{\alpha\beta}{}_{\rho\sigma}F^{\rho\sigma}/2=v^\alpha B^\beta-v^\beta B^\alpha+\eta^{\alpha\beta}{}_{\rho\sigma}v^\rho E^\sigma$ stands for the dual of the Maxwell field strength tensor $F^{\alpha\beta}=v^\alpha E^\beta-v^\beta E^\alpha-\eta^{\alpha\beta}{}_{\rho\sigma}v^\rho B^\sigma$, while the Faraday tensor $P^{\alpha\beta}=\varepsilon(v^\alpha E^\beta-v^\beta E^\alpha)-(1/\mu)\eta^{\alpha\beta}{}_{\rho\sigma}v^\rho B^\sigma$ describes the field excitations. We assume that the permittivity parameter $\epsilon=\epsilon(E)$ may be dependent upon the magnitude of the electric field strength $E\equiv\sqrt{-E_{\alpha}E^{\alpha}}$, while the permeability parameter is momentarily being taken as the vacuum constant $\mu=\mu_0$.

The electromagnetic field strengths $E^\alpha$ and $B^\alpha$ are assumed both to be continuous but with possibly non-zero finite Hadamard discontinuities \cite{hadamard} in their derivatives at the wave-front hypersurface $\Sigma$, as
\begin{eqnarray}
&\left[E_{\alpha,\beta}\right]_{\Sigma}=e_\alpha k_\beta
\label{e},\\[1ex]
&\left[B_{\alpha,\beta}\right]_{\Sigma}=b_\alpha k_\beta
\label{b},
\end{eqnarray}
where $k_\beta=\partial_\beta\Phi=(\omega,\vec q)=q(v_{ph},\hat q)$ is the wave vector (with a phase speed $v_{ph}$ pointing along the direction of the normalized vector $\hat q$) orthogonal to $\Sigma$, while $\Sigma$ is described as $\Phi(x^\alpha)=0$. The two space-like vectors $e^\alpha$ and $b^\alpha$ respectively describe the polarizations of the electric and magnetic components of the wave. The discontinuity of Eqs.\ (\ref{max_diel1})--(\ref{max_diel2}) over $\Sigma$ yields a linearly polarized wave $b^\alpha=\eta^{\alpha\beta}{}_{\rho\sigma}k_\beta v^\rho e^\sigma/\omega$, with the electric polarization $e^\alpha$ being obtained from the eigenvalue problem $Z^\alpha{}_\beta e^\beta=0$, where the generalized Fresnel matrix is written in terms of the normalized vector $\hat l{}^\alpha=E^\alpha/E$ as
\begin{equation}
Z^\alpha{}_\beta=\epsilon(\delta^\alpha_\beta-v^\alpha v_\beta)-\epsilon'E\hat l^\alpha\hat l_\beta-\frac1{\mu v_{ph}^2}(\delta^\alpha_\beta-v^\alpha v_\beta+\hat q{}^\alpha\hat q{}_\beta),
\label{Z}
\end{equation}
where $\epsilon'$ denotes the derivative of the function $\epsilon(E)$ with respect to $E$.
This relation was written in terms of $\mu$ (instead of $\mu_0$) for latter convenience. The existence of non-trivial eigenvectors $e^\alpha\neq0$ then leads \cite{novello5,gordon,hehl} to the two possible optical geometries
\begin{eqnarray}
&g_{(+)}^{\alpha\beta}=\gamma^{\alpha\beta}-(1-\mu\epsilon)v^\alpha v^\beta
\label{eff_met+},\\[2ex]
&g_{(-)}^{\alpha\beta}=\gamma^{\alpha\beta}-[1-\mu(\epsilon+\epsilon'E)]v^\alpha v^\beta-\fracc{\epsilon'E}{\epsilon}\hat l{}^\alpha\hat l{}^\beta
\label{eff_met-}\label{opt_up},
\end{eqnarray}
such that the two wave vectors $k_\lambda=k_\lambda^{(\pm)}$ satisfy the corresponding dispersion relations $g_{(+)}^{\alpha\beta}k_\alpha^{(+)}k_\beta^{(+)}=0$ and $g_{(-)}^{\alpha\beta}k_\alpha^{(-)}k_\beta^{(-)}=0$. The former equation is equivalent to $\mu\epsilon\omega_{(+)}^2=q_{(+)}^2$, which describes the isotropic propagation of the {\em ordinary} mode with speed $v_{ph}=1/\sqrt{\mu\epsilon}$. The effective optical geometry Eq.\ (\ref{eff_met+}) is usually referred to as Gordon geometry .

Our interest here lies mostly in the so-called {\em extraordinary} mode, the optical geometry Eq.\ (\ref{eff_met-}), and thus we simply denote it as $g^{\alpha\beta}=g_{(-)}^{\alpha\beta}$. Assuming that the determinant of $g^{\alpha\beta}$ is non zero, then a simple calculation \cite{novello5} gives its inverse matrix as the effective metric
\begin{equation}
g_{\alpha\beta}=\gamma_{\alpha\beta}-\left[1-\fracc{1}{\mu(\epsilon+\epsilon'E)}\right]v_\alpha v_\beta+\fracc{\epsilon'E}{\epsilon+\epsilon'E}\hat l{}_\alpha\hat l{}_\beta
\label{opt_down}.
\end{equation}

The results above can easily be generalized to include the case for which also the permeability parameter has an arbitrary dependence upon the electric field strength as $\mu=\mu(E)$. Indeed, the Fresnel matrix Eq.\ (\ref{Z}) and the two optical geometries Eqs.\ (\ref{eff_met+})--(\ref{eff_met-}) can be shown \cite{goulart} to hold good in this case as well, provided the magnetic field $\vec B$ is zero. We will henceforth assume this condition to be satisfied. Let us now seek for static spherically symmetric black hole analogue models to this effective optical metric.

\section{Analogue spherical black holes}
\label{Analogue}
Let us consider a dielectric medium as above, with four-velocity $v^{\alpha}=\delta^{\alpha}_0$, subjected to an electric field directed along the radial direction and no magnetic field. For the static spherically symmetric situation we are dealing with, the current four-vector $J^\mu=(\rho,\vec J)$ presents only its time component, the charge density $\rho$. Maxwell Eqs.\ (\ref{max_diel1})--(\ref{max_diel2}), written in flat spherical coordinates $(t,r,\theta,\phi)$ adapted to the dielectric medium, then reduce to
\begin{equation}
\frac{\partial_r(\sqrt{-\gamma}\epsilon E)}{\sqrt{-\gamma}}=\rho
\label{max},
\end{equation}
where the determinant of the metric $g_{\alpha\beta}$ is $\gamma=-r^4\sin^2\theta$ as in the flat case. The effective geometry Eq.\ (\ref{opt_up}) reads
\begin{equation}
g^{\alpha\beta}={\rm diag}\left(\mu(\epsilon+\epsilon'E),\,-\frac{\epsilon+\epsilon'E}{\epsilon},\,-\frac{1}{r^2},\,-\frac{1}{r^2\sin^2\theta}\right)
\label{diag}.
\end{equation}
This form allows one to seek for analogue spherically symmetric static black hole solutions
\begin{equation}
g^{\alpha\beta}={\rm diag}\left(\frac{1}{A},\,-A,\,-\frac{1}{r^2},\,-\frac{1}{r^2\sin^2\theta}\right)
\label{bh},
\end{equation}
where $A=A(r)$ is a given radial function such that $A(r)=1-R/r$ describes a Schwarzschild black hole with a horizon at the Schwarzschild radius $R$. Equation (\ref{bh}) includes all spherically symmetric black holes in a unified form, with the well-known solutions of Einstein equations (such as Schwarzschild, Reissner-Nordstr\"om, de-Sitter or combinations thereof) being characterized by the explicit form of the radial dependence of the function $A$. The identification of Eqs.\ (\ref{diag}) and (\ref{bh}) gives the two possible solutions
\begin{equation}
\epsilon+\epsilon'E=\pm\sqrt{\epsilon/\mu}
\label{eq_epsilon}.
\end{equation}
In order to integrate this equation, a two-parameter function
$\mu=\mu(\epsilon,E)$ can arbitrarily be chosen. Among all possibilities, we restrict ourselves here (other possible choices are left to the concluding discussions) to the mathematically convenient form
\begin{equation}
\mu=\frac{\epsilon_0^2}{\epsilon^3}
\label{mu},
\end{equation}
where $\epsilon_0$ is the vacuum permittivity constant (with $\mu_0\epsilon_0=1$). Equation (\ref{eq_epsilon}) then reads $(\epsilon E)'=\pm\epsilon^2/\epsilon_0$, whose integration immediately yields $\epsilon=\epsilon_\pm$, where
\begin{equation}
\epsilon_\pm=\fracc{\epsilon_0}{\frac{E}{E_0}\pm1}
\label{epsilon},
\end{equation}
and $A(r)=\pm\epsilon_\pm/\epsilon_0=1/(1\pm E/E_0)$, where $E_0>0$ is a constant of integration; for $\epsilon=\epsilon_{{}+{}}$ one has $E=E_0$ at $A=1/2$ ({\em i.e.}, at $r=2R$ for the Schwarzschild model), while the solution $\epsilon=\epsilon_{{}-{}}$ is limited to $E>E_0$ (since $E<E_0$ would correspond to $A>1$ in this case). The usual range $-\infty<A<1$ of an effective black hole can thus be obtained by joining the solution $\epsilon_{{}-{}}$ inside horizon with the solution $\epsilon_{{}+{}}$ outside horizon.

The expressions of the electric field $E$, the electric displacement $D=\epsilon E$, and the charge density $\rho$ in terms of $A$ then give
\begin{eqnarray}
&\fracc{E}{E_0}=\pm\fracc{(1-A)}{A}
\label{E-A},\\
&D=\epsilon_0E_0(1-A)
\label{D-A},\\
&\rho=\epsilon_0E_0\left[\fracc{2(1-A)}{r}-\fracc{{\rm d}A}{{\rm d}r}\right]
\label{rho},
\end{eqnarray}
which hold for either $0<A<1$ or $A<0$. In the case of a Schwarzschild analogue black hole, these expressions reduce to $E=\pm E_0R/(r-R)$ with a quadratic charge density profile $\rho=\epsilon_0E_0R/r^2$ and a linear electric displacement $D=\epsilon_0E_0R/r$ (note that $\rho/D=1/r$ in this case). Therefore, $E$ diverges at the horizon but remains finite everywhere else, while both $D$ and $\rho$ are finite at the horizon but they both diverge at the center (except for $A-1\sim r^{-2}$, which gives $\rho=0$ everywhere). The inner solution should then be regularized near the center. Our definite proposal to a spherically symmetric black hole analogue with radius $R$ is thus built with a medium whose permittivity is such that
\begin{equation}
\frac{\epsilon}{\epsilon_0}=\left\{
\begin{array}{ll}
\frac{E_0}{E+E_0},&\mbox{if }r>R,\\
\frac{E_0}{E-E_0},&\mbox{if }\frac12R<r<R,\\
1,&\mbox{if }r<\frac12R.
\end{array}
\right.
\label{eps(r)}
\end{equation}
When expressed in terms of the radial coordinate $r$, then Eq.\ (\ref{eps(r)}) gives $\epsilon/\epsilon_0=|A|$, where $-1<A<1$. We can compare this result with previous similar proposals for the radial behavior of a nonlinear dielectric medium at rest: for example,  \cite{narimanov} which rely upon postulating a core absorption coefficient; here, no doping is required, but only a variable volumetric density of the medium. Moreover, as already noted \cite{chen}, that was not a consistent solution of Einstein field equations; this latter instead deals with the cylindric case, and proposed a non-diagonal structure for both $\epsilon$ and $\mu$, while we treat these two parameters both as scalars.
\begin{figure}[!hbt]
\leavevmode
\centering
\includegraphics[scale = .65]{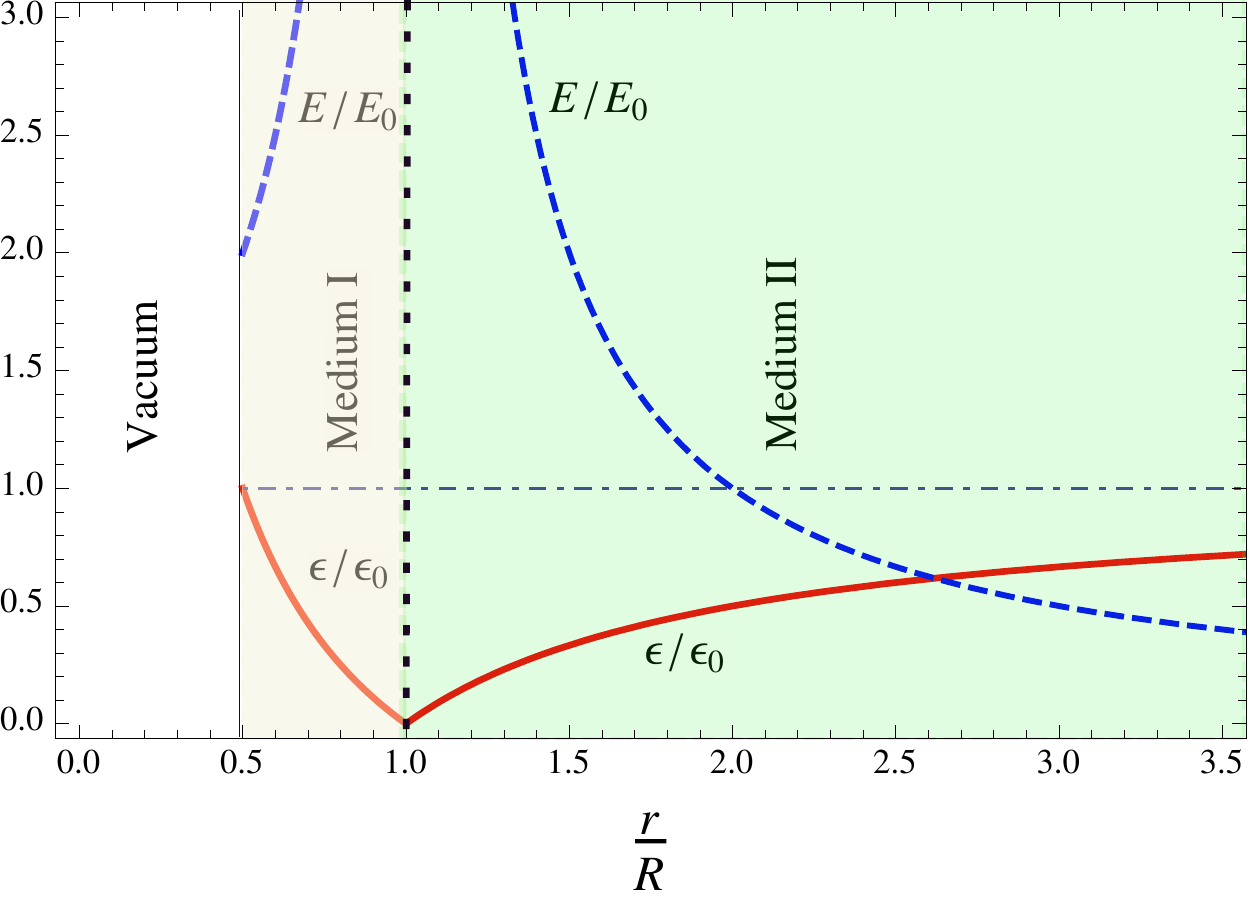}
\caption{{\small\sf (color online). The behavior of the dielectric coefficient $\epsilon$ and the electric field $E$ are shown in terms of the radial distance $r$. As stated by Eq.~(\ref{eps(r)}), three distinct media are considered in this model: vacuum for $0<r<0.5R$, medium I for $0.5<r<R$ and medium II for $r > R$. Notice that $\epsilon$ is a $C^0$ function at the horizon, while the electric field diverges at this point. However, the observable field $D$ is a regular function through $r=R$. The dotted vertical line depicts the analogue event horizon.}}
\label{fig1}
\end{figure}

\section{Kinematics}
\label{Kinematics}
The usefulness of the optical geometry Eq.\ (\ref{bh}) is to express the dispersion relation as $g^{\alpha\beta}k_\alpha k_\beta=0$. Written in terms of the phase speed $v_{ph}=\omega/q$ of the wave, we have
$k_\lambda=\omega v_\lambda+q_\lambda=q\,(v_{ph}v_\lambda+\hat q{}_\lambda)$,
and then
\begin{equation}
v_{ph}^2=A[1+(A-1)(\hat l\cdot\hat q)^2]
\label{vph}.
\end{equation}
The phase velocity $\vec v{}_{ph}=v_{ph}\hat q$ and the group velocity $\vec v{}_{gr}={\rm d}\omega/{\rm d}\vec q$ are
\begin{eqnarray}
&&\vec v{}_{ph}=|A|^{1\!/\!2}|1+(A-1)(\hat l\cdot\hat q)^2|^{1\!/\!2}\hat q
\label{phase},\\
&&\vec v{}_{gr}=\sqrt{\fracc{|A|}{|1+(A-1)(\hat l\cdot\hat q)^2|}}\,[\hat q+(A-1)(\hat l\cdot\hat q)\hat l].\nonumber\\
&&\label{group}
\end{eqnarray}
These two velocities coincide for each of the following three situations: $\hat l\cdot\hat q=1$, $\hat l\cdot\hat q=0$, or $\hat l\cdot\hat q=-1$. For other directions $\hat q$ of the phase velocity with respect to the radial direction $\hat l$ of the background electric field $\vec E$, the two velocities $\vec v{}_{ph}$ and $\vec v{}_{gr}$ given by Eqs.\ (\ref{phase})--(\ref{group}) lie along lines not parallel to one another. Let $\psi$ be the angle between $\vec v{}_{gr}$ and $\vec v{}_{ph}$. It then follows from Eqs.\ (\ref{phase})--(\ref{group}) that
\begin{equation}
(\hat l\cdot\hat q)^2=\frac{1-\cos\psi}{1-A}
\label{psi},
\end{equation}
which implies $\cos\psi\ge A$. In particular, $\cos\psi>0$ outside the effective black hole ({i.e.}, for the region $0<A<1$).

It should be remarked that both the phase and group velocities of this extraordinary mode have a zero limit when approaching the regular distinguished closed surface $A=0$, lacking propagation of this wave mode across that surface.

For the {\em ordinary} mode, and taking into account the solution Eq.\ (\ref{epsilon}) for $\epsilon$, we obtain
\begin{equation}
v_{ph}^{(+)}=|A|=v_{gr}^{(+)}
\label{v+},
\end{equation}
where $A$ is the same radial function considered above in Eq.\ (\ref{bh}). Equation (\ref{v+}) states that the {\em ordinary} mode $k_\lambda^{(+)}$ also cannot propagate across the surface $A=0$. Therefore, this surface plays the role of the effective event horizon of the above considered optical analogue model of a black hole.

The arbitrary constant $E_0$ can be eliminated from Eqs.\ (\ref{D-A})--(\ref{rho}), thus yielding ${\rm d}A/{\rm d}r=(1-A)(2/r-\rho/D)$, from which the standard definition of the surface gravity parameter $\kappa=(c^2/2)\lim_{r\to R}{\rm d}A/{\rm d}r$ reads
\begin{equation}
\kappa=\lim_{r\to R}\left(\frac{\rho}{2D}\right)
\label{kappa}.
\end{equation}
Particularization of this result to the Schwarzschild model gives $\kappa=[\rho/(2D)]|_{r=R}=1/(2R)$.

In order to have some estimates, suppose a dielectric probe with an optical Schwarzschild radius of $R=10\,\mbox{cm}$.  The total amount of electric charge in the volume delimited by the radius $R/2$ and $2R$ of the dielectric would be of the order $Q\sim10^{-6}C$, while the electric field scale $E_0$ is given by $E_0=Q/6\pi\epsilon_0R^2\approx6.0\cdot 10^5V/\mbox m$. The charge density in this region of radial distances thus ranges between $\rho_{min}=\epsilon_0E_0/4R\approx1.33\cdot 10^{-5}C/\mbox m^3$ and $\rho_{max}=4\epsilon_0E_0/R\approx2.12\cdot 10^{-4}C/\mbox m^3$.
For the sake of comparison, the above values are similar to the ones we can find in usual electronic capacitors with capacitance ranging around $(1-50)\mu F$ and providing an electrostatic potential of $1\rm V$.
From the regularity of the physical properties mentioned above, we see that the proposed dielectric media can reproduce in laboratory all the classical behavior of optical null geodesics near a compact object with a gravitational Schwarzschild radius equal to $R$.

\section{Conclusion}
\label{Conclusion}
Suppose a static spherically symmetric dielectric medium with electric charge density $\rho$ given by Eq.\ (\ref{rho}), while its dielectric parameters $\epsilon$ and $\mu$ are real quantities (that is, with no `{by-hand}' absorption) which behave non-linearly according with Eqs.\ (\ref{mu}) and (\ref{eps(r)}). Maxwell equations then yield no magnetic field and a radial electric field given by Eq.\ (\ref{E-A}). With such electromagnetic background fields, small electromagnetic field disturbances propagate as two linearly polarized wave modes. The extraordinary one amongst these two modes behaves exactly as a null wave of the black hole analogue geometry Eq.\ (\ref{bh}). The ordinary mode propagates differently, but in such a way as to avoid propagation across the same analogue event horizon. Thus, no electromagnetic field disturbance can propagate from the inner region of the analogue black hole to the region outside from it.

As well known, quantization of fields on the classical Schwarzschild spacetime (semiclassical gravity) leads to the concept of Hawking radiation phenomenon. In terms of the surface gravity $\kappa$, the Planckian spectrum associated with such radiation defines the temperature $T_H=\hbar c\kappa/(2\pi k_B)$. For the case of astrophysical candidates of black-holes the magnitude of this temperature is too small to be experimentally probed. Analogue models of the Schwarzschild solution could thus provide a useful arena to investigate Hawking radiation. For instance, if we naively accept that Hawking radiation is produced in a system described by the model examined in the last section, we would obtain $T_H \approx1.822\cdot 10^{-4}[K\cdot\mbox m]/R$, which lies in the accessible range scale of conventional thermometers for a material sample with a few centimeters in size.
However, a proper demonstration of the existence of this phenomenon requires the quantization of the electromagnetic field in the nonlinear media described by the dielectric coefficients here proposed. This is an issue that deserves a careful analysis.

As an apparently simpler alternative to Eq.\ (\ref{mu}), the choice $\mu=\mu_0$ could also be worked out as well, yielding $E=E_0A^2/(1-A)^2$. For such a case, however, the background electric field would vanish at horizon $A=0$. Thus, the small magnitude of the electric field of the wave could not be neglected in Eq.\ (\ref{opt_up}) when compared to the background electric field. This means that Eq.\ (\ref{opt_up}) would not be an effective geometry very near the horizon, but it would instead depend as well on the field of the propagating wave there. That is to say, the analogue event horizon would become rather blurred and undefined in such simple case.

Other choices of the form $\mu=\epsilon_0^n/\epsilon^{n+1}$ with arbitrary constant $n\neq0$ yield qualitatively similar results as the ones presented in the preceding sections for $n>0$, or those mentioned in the previous paragraph for $n<0$; such a class includes the proportionality condition \cite{visser2011} between $\epsilon$ and $\mu$ claimed to be required in order to give room for the optical equivalence beyond the geometrical limit. Other alternative choices for $\mu(\epsilon,E)$ may possibly prove useful as well to help understanding better optical analogue models.


\section*{Acknowledgements}
E. B. would like to thank Universidade Federal de Itajub\'a for the hospitality.
This work was partially supported by CNPq, CAPES (BEX 18011/12-8 and 13956/13-2), FINEP and FAPERJ.


\begin{thebibliography}{88}
\bibitem{gordon}
W. Gordon, {\em Ann.\ Phys.\ (Leipzig)} {\bf 72}, 421  (1923).
\bibitem{plebanski1960}
J. Plebanski, {\em Phys.\ Rev.\ } {\bf 118}, 1396 (1960).
\bibitem{landau1971}
L. D. Landau and E. M. Lifshitz, {\em The classical theory of fields} (Pergamon Press Ltd., Oxford, 1971), p 256.
\bibitem{felice1971}
F. de Felice, {\em Gen.\ Rel.\ Grav.\ } {\bf 2}, 347 (1971).
\bibitem{schenberg1971}
M. Schenberg, {\em Revista Brasileira de Fisica}, {\bf 1}, 91 (1971).
\bibitem{unruh1981}
W. G. Unruh, {\em Phys.\ Rev.\ Lett.\ } {\bf 46}, 1351 (1981).
\bibitem{visser2011}
C. Barcel\'o, S. Liberati, and M. Visser, {\em Living Rev.\ Rel.\ } {\bf 8}, 12 (2005);
ibidem {\bf 14}, 3 (2011).
\bibitem{belinski}
V. A. Belinski, {\em Phys.\ Lett.\ A} {\bf 209}, 13 (1995).
\bibitem{hawking1975}
S. W. Hawking, {\em Commun.\ Math.\ Phys.\ } {\bf 43}, 199 (1975).
\bibitem{belgiorno2010}
F. Belgiorno {\it et al.}, {\em Phys.\ Rev.\ Lett.} {\bf 105}, 203901 (2010).
\bibitem{unruh2011}
S. Weinfurtner, E. W. Tedford, M. C. J. Penrice, W. G. Unruh, and G. A. Lawrence, {\em Phys.\ Rev.\ Lett.\ } {\bf 106}, 021302 (2011).
\bibitem{edu2012}
M. Novello and E. Bittencourt, {\em Phys.\ Rev.\ D} {\bf 86}, 124024 (2012).
\bibitem{novello4}
M. Novello, S. E. Perez-Bergliaffa, J. M. Salim, V. A. De Lorenci, and R. Klippert, {\em Class.\ Quantum Grav.\ } {\bf 20}, 859 (2003).
\bibitem{Yuri}
V. A. De Lorenci, R. Klippert, and Yu.\ N. Obukhov, {\em Phys.\ Rev.\ D} {\bf 68}, 061508R (2003).
\bibitem{bjp}
V. A. De Lorenci and R. Klippert, {\em Braz.\ J. Phys.\ } {\bf 34}, 1367 (2004).
\bibitem{narimanov}
E. E. Narimanov and A. V. Kildishev, {\em Appl.\ Phys.\ Lett.\ } {\bf 95}, 041106 (2009).
\bibitem{sheng}
C. Sheng, H. Liu, Y. Wang, S. N. Zhou, and D. A. Genov, {\em Nature Photonics} {\bf 7}, 902-906 (2013).
\bibitem{yin}
M. Yin, X. Y. Tian, L. L. Wu, and D. C. Li, {\em Opt.\ Exp.} {\bf 21} 19082 (2013).
\bibitem{local}
V. A. De Lorenci and R. Klippert, {\em Phys.\ Lett.\ A} {\bf 357}, 61 (2006).
\bibitem{hadamard}
J. Hadamard, \textit{Le\c cons sur la propagation des ondes et les \'equations de hydrodynamique} (Hermann, Paris, 1903);
V. D. Zakharov, \textit{Gravitational waves in Einstein's theory} (John Wiley \& Sons, Inc., New York, 1973).
\bibitem{hehl}
F. W. Hehl and Y. N. Obukhov, \textit{Foundations of classical electrodynamics: charge, flux, and metric}, Progress in Mathematical Physics, v.\ 33 (Birkh\"auser, Boston, 2003).
\bibitem{novello5}
M. Novello and J. M. Salim {\em Phys.\ Rev.\ D} {\bf 63}, 083511 (2001).
\bibitem{goulart}
V. A. De Lorenci and G. P. Goulart, {\em Phys.\ Rev.\ D} {\bf 78}, 045015 (2008).
\bibitem{chen}
H.-Y. Chen, R.-X. Miao, and M. Li, {\em Opt.\ Express} {\bf 18}, 15183-15188 (2010).
\end{thebibliography}
\end{document}